\documentclass[aps,prl,superscriptaddress,longbibliography,twocolumn]{revtex4-2}
\usepackage{epsf,graphicx}
\usepackage{amssymb}
\usepackage{amsmath}
\usepackage{latexsym,bm,array,amsfonts,multirow}
\usepackage{color}
\usepackage{ulem}
\usepackage{epstopdf}
\begin{document}
	
\title{Symmetry-resolved superconducting diode response
}
\author{Qiong Qin}
\affiliation{New Cornerstone Science Laboratory, Department of Physics, Westlake University, Hangzhou 310024, Zhejiang, China}
\affiliation{Department of Physics, Westlake University, Hangzhou 310024, Zhejiang, China}
\author{Jie Wu}
\affiliation{Department of Physics, Westlake University, Hangzhou 310024, Zhejiang, China}
\affiliation{Key Laboratory for Quantum Materials of Zhejiang Province, School of Science, Westlake University, Hangzhou 310024, Zhejiang, China}
\affiliation{Institute of Natural Sciences, Westlake Institute for Advanced Study, Hangzhou 310024, Zhejiang, China}
\author{Congjun Wu}
\email[]{wucongjun@westlake.edu.cn}
\affiliation{New Cornerstone Science Laboratory, Department of Physics,  Westlake University, Hangzhou 310024, Zhejiang, China}
\affiliation{Institute for Theoretical Sciences, Westlake University, Hangzhou 310024, Zhejiang, China}
\affiliation{Key Laboratory for Quantum Materials of Zhejiang Province, School of Science, Westlake University, Hangzhou 310024, Zhejiang, China}
\affiliation{Institute of Natural Sciences, Westlake Institute for Advanced Study, Hangzhou 310024, Zhejiang, China}

\begin{abstract}
The superconducting diode effect manifests as nonreciprocal properties of critical currents, yet a systematic tensorial framework for its symmetry-governed responses remains lacking. 
We develop such a framework by relating the angular harmonics of critical currents to the magnetic field.
Crystalline symmetries constrain the dipolar response to be purely antisymmetric (Hall-like) in Rashba superconductors with $C_{3v}$, $C_{4v}$, or $C_{6v}$ symmetry, whereas they require a diagonal, traceless tensor in Dresselhaus superconductors with $D_{2d}$ symmetry.
For systems with fourfold symmetry, the octupole response appears at linear order, whose response tensors take symmetric off-diagonal and identity forms for $C_{4v}$ and $D_{2d}$ symmetries, respectively.
Furthermore, nematicity activates otherwise forbidden tensor components, providing a sensitive probe of rotational symmetry breaking in superconductors. 
At cubic order in the magnetic field, the nonlinear response tensors likewise exhibit distinct symmetry-dependent patterns.
Our framework establishes a symmetry-resolved characterization of the superconducting diode effect and provides a route to probing the underlying electronic structure and symmetry-breaking orders.
\end{abstract}
\maketitle

\textit{Introduction.--}
The superconducting (Josephson) diode effects are characterized by nonreciprocal critical currents, manifested as different critical-current magnitudes for opposite bias voltages in the AC effect \cite{Hu2007} and for opposite current directions in the DC effect \cite{Wakatsuki2018,Jiang2022}.
It has been experimentally observed in a broad range of superconducting systems, both with and without Josephson junctions, and in the presence or absence of external magnetic fields
\cite{Ando2020a,Narita2022,Lin2022b,Pal2022,Nagata2025,Hou2023a,
Le2024a,Qi2025,Anh2024,Valentini2024,Wang2026,Nadeem2023,
Li2024a,Kudriashov2025,Margineda2023,Margineda2026}.
These observations have stimulated extensive interest in identifying the microscopic mechanisms responsible for the diode effect  
\cite{Daido2022a,Daido2022b,He2022a,Yuan2022a,Scammell2022,
Ilic2022,Mao2024a,Banerjee2024,Lu2023a,Hu2023,Zhang2022d,
Wang2022c,Wang2026arxiv,Varma2025,Wang2025arxiv,Hu2025,
Misaki2021,Fracassi2024}.

Despite extensive studies of the superconducting diode effect, a global, symmetry-based characterization of the nonreciprocal transport properties remains lacking.
While the commonly used diode efficiency quantifies nonreciprocity along a given direction, it does not by itself capture the full angular response, which encodes information determined by the underlying symmetries of the system.
A global, symmetry-sensitive description of the superconducting diode effect would make it a powerful probe of superconducting states beyond conventional transport measurements.



Here, we propose to use the angular harmonics, {\it i.e.}, multipole components of the angular distribution of critical current, to describe the nonreciprocity of a superconducting system.
The multipole components are related to an applied in-plane magnetic field via the response tensors that are determined by crystalline symmetries.
This framework establishes a symmetry-based description of response tensors in different angular harmonic channels.
It thus offers a systematic way to characterize the superconducting diode effect beyond a direction-dependent scalar efficiency.


In this article, we find that in systems with the crystalline symmetries of $C_{3v}$, $C_{4v}$, or $C_{6v}$, 
as exemplified by systems of Rashba spin-orbit (SO) coupling \cite{Bychkov1984,Manchon2015a}, 
the dipolar response tensor at the linear order takes an antisymmetric form.
The presence of nematicity, however, breaks rotational symmetry and induces a symmetric component. 
In contrast, systems dominated by the Dresselhaus SO coupling \cite{Dresselhaus1955,Manchon2015a} with the $D_{2d}$ symmetries yield a diagonal traceless tensor, while nematicity further generates a finite trace component.
In systems with 4-fold symmetries (e.g. $C_{4v}$ and $D_{2d}$), octupole responses are also generated at the linear level.
We extend the symmetry analysis to cubic order in End Matter.
These response tensors can be directly extracted from measurements of the superconducting diode effect, which offers a promising and accessible route to probe nematic order \cite{Xiang2021a,Hamill2021,Sigrist2005} within the superconducting phase.

\textit{Response tensors to superconducting diode effect.--} 
Consider the angular dependence of the critical current density $J_c(\phi)$ in a 2D superconducting state, where $\phi$ is the azimuthal angle of the 
supercurrent transport.
In literature, the non-reciprocal relation is often expressed as $\Delta J_c (\phi) = J_c(\phi) - J_c(\phi + \pi)$.
To facilitate a comprehensive characterization of the superconducting diode effect, we propose to employ the multipolar components of 
$J_c (\phi)$ in the odd-partial wave channels.
In 2D, they are
$\bm{D}^{(n)} = 
(D_{1}^{(n)}, D_{2}^{(n)})$ 
defined as 
\begin{eqnarray}
D_{1}^{(n)} (D_{2}^{(n)})&=&\frac{1}{2}\int_0^{2\pi} d\phi~ \Delta J_c(\phi) \cos n\phi~ (\sin n\phi),
\label{eq:D1}
\end{eqnarray} 
where $n=2k-1$ is an odd number. 

We investigate the non-reciprocal response of $\bm{D}^{(n)}$ to an in-plane magnetic field $\bm{B}$ which breaks time-reversal symmetry.
In a 2D lattice system, only discrete rotational symmetry is kept.
For simplicity, we consider the case of 4-fold symmetry in which the angular momentum is conserved modulo 4. 
Hence, even at the linear level of response, the in-plane $\bm{B}$  generates all the odd partial-wave multipole components of $\bm{D}^{(n)}$.
For simplicity, we consider two lowest components of dipole $\bm{D}^{(1)}$ and octupole $\bm{D}^{(3)}$. 
The rank-2 response tensors are defined
\begin{eqnarray}
D_i^{(n)}=\chi^{(n)}_{ij} B_j,  
\label{eq:linear_resp}
\end{eqnarray}
where $n=1$ and $3$, respectively.
$\hat{\bm{\chi}}^{(n)}$ carries the same unit as electrical conductance in the Guassian system.
In the absence of the diode effect, $\Delta J_c(\phi) = 0$, leading to $\bm{D}^{(n)} = \bm{0}$ and a vanishing response tensor $\hat{\bm{\chi}}^{(n)}$.
A nonzero $\hat{\bm{\chi}}^{(n)}$ therefore signals the superconducting diode effect. 


The symmetry properties impose stringent  constraints on the response tensors \cite{He2020}.
Symmetries that reverse the current direction are dubbed the current reversal (JR) symmetries  \cite{Wang2022c}.
Two most frequently studied JR symmetries are inversion $\mathcal{P}$ and time-reversal $\mathcal{T}$. (In our context ${\cal T}$ refers to the symmetry 
of the system
in the absence of external B-field.)
The critical current density transforms as $J_c(\phi) \rightarrow J_c(\phi + \pi)$, hence $\bm{D}^{(n)}$ flips the sign.
In contrast, $\bm{B}$ is even under $\mathcal{P}$ and odd under $\mathcal{T}$, respectively. 
Consequently $\hat{\bm{\chi}}^{(n)}$ is odd under $\mathcal{P}$  and even under $\mathcal{T}$.
Breaking ${\cal P}$ is therefore a necessary condition for the superconducting diode effect.
As for a point-group operation $O$, 
$\bm D^{(n)}$ transforms as 
$D_i^{\prime (n)}=\sum_{i'}R^{(n)}_{ii'}[O] D_{i'}^{(n)}$, 
and $\bm B$ transforms as a pseudo-vector, 
$B_j^{\prime}=	{\rm sign}[O]\sum_{j'}R^{(1)}_{jj'}[O]B_{j'}$,
where ${\rm sign}[O]=\pm1$ for proper and improper point group operations, respectively. 
Then we arrive at 
\begin{eqnarray}
\hat{\bm{\chi}}^{(n)} = {\rm sign}[O] 
R^{(n),T}[O]~ \hat{\bm{\chi}}^{(n)}~ R^{(1)}[O],
\end{eqnarray}
where $T$ means the matrix transpose.


We decompose $\hat{\bm{\chi}}^{(n)}$ in terms of the basis of the Pauli matrices: $\hat{\bm{\chi}}^{(n)}=\chi_0^{(1)} I +
\chi_1^{(1)} \tau_1 +\chi_3^{(1)} \tau_3 +\chi_2^{(1)} i\tau_2$, where $I, \tau_{1,3}$ are symmetric and $i\tau_2$ is anti-symmetric.
Consider the dipolar response 
$\bm{\chi}^{(1)}$ for a system under 
$C_{4v}$ symmetry. 
Check the mirror symmetry with respect the $xz$-plane, {\it i.e.}, $M_{xz}$,
which is
improper and its vector representation is given by $R^{(1)}[M_{xz}]=\tau_3$.
Then $\tau_3 \hat{\bm{\chi}}^{(1)} \tau_3 =-\hat{\bm{\chi}}^{(1)}$, 
giving rise to $\chi_0^{(1)}=\chi_3^{(1)}=0$. 
Furthermore, the 4-fold rotational symmetry $C_{4z}$ is represented by 
$R^{(1)}[C_{4z}]=-i\tau_2$, 
then $\tau_2 \hat{\bm{\chi}}^{(1)} \tau_2 =\hat{\bm{\chi}}^{(1)}$, leading to
$\chi_1^{(1)}=\chi_3^{(1)}=0$. 
Hence, the $C_{4v}$ symmetry
yields a fully antisymmetric tensor
$\bm{\hat{\chi}}^{(1)}=\chi_2^{(1)} i\tau_2$,
and the associated 
first harmonic angular distribution of $\Delta J^{(1)}_c(\phi)$ as
\begin{eqnarray}
\Delta J^{(1)}_c(\phi)=
\frac{2}{\pi}
B\chi_2^{(1)} \sin (\phi_B-\phi),
\label{eq:c4v_dip}
\end{eqnarray}
where $\phi_B$ is the azimuthal angle of the in-plane B-field.
In addition, 
by similar reasoning, 
for systems with the $C_{3v}$ and $C_{6v}$ symmetries, their dipolar responses to the in-plane B-field are the same as $C_{4v}$. 


For a system with the $D_{2d}$ symmetry, the $S_4$ and  $C_{2x}$ replace the roles of 
$C_{4z}$ and $M_{xz}$, respectively.
The $S_4$ operation, {\it i.e.}, a 4-fold rotation followed by a reflection with respect to the $xy$-plane, is improper
and its vector representation is 
$R^{(1)}[S_4]=-i\tau_2$. 
It leads to $\tau_2
\hat{\bm{\chi}}^{(1)} \tau_2 =- \hat{\bm{\chi}}^{(1)}$, and then $\chi_0^{(1)}=\chi_2^{(1)}=0$. 
As for $C_{2x}$, whose vector representation is $R^{(1)}[C_{2x}]=\tau_3$, hence, 
$\tau_3 \hat{\bm{\chi}}^{(1)}
\tau_3= \hat{\bm{\chi}}^{(1)}$, leading to 
$\chi_1^{(1)}=\chi_2^{(1)}=0$.
Therefore, when both $S_4$ and $C_{2x}$ (or $C_{2y}$) are present, we arrive at $\bm{\hat{\chi}}^{(1)}=\chi_3^{(1)} \tau_3$, and
\begin{eqnarray}
\Delta J^{(1)}_c(\phi)=\frac{2}{\pi} B\chi_3^{(1)} \cos (\phi+\phi_B).
\label{eq:d2d_dip}
\end{eqnarray}

Next we consider the response tensor $\bm{\hat \chi}^{(3)}$ of the octupolar component $\bm D^{(3)}$ to the in-plane magnetic field. 
The point group operation $O$ transforms 
$\bm{D}^{(3)}$ in an equivalent way to 
$\bm{D}^{(1)}$ through $R^{(3)}[O]
=\tau_3 R^{(1)}[O] \tau_3$.
Now study the constraint from the $C_{4v}$ symmetry:
The mirror symmetry $M_{zx}$ is represented by $R^{(3)}[M_{xz}] =\tau_3$, hence,  
$\tau_3 \hat{\bm{\chi}}^{(3)} \tau_3 =-\hat{\bm{\chi}}^{(3)}$.
$C_{4z}$ is represented by 
$R^{(3)}[C_{4z}]=-R^{(1)}[C_{4z}]=i\tau_2$.
Hence, $\tau_2 \hat{\bm{\chi}}^{(3)}
\tau_2= -\hat{\bm{\chi}}^{(3)}$.
Hence, the $C_{4v}$ symmetry constrains 
$\bm{\hat{\chi}}^{(3)}=\chi_1^{(3)} \tau_1$,
and the associated third harmonic angular distribution of $\Delta J^{(3)}_c(\phi)$
as
\begin{eqnarray}
\Delta J^{(3)}_c(\phi)=\frac{2}{\pi} B\chi_1^{(3)} \sin (3\phi+\phi_B).
\label{eq:c4v_oct}
\end{eqnarray}
As for the $D_{2d}$ symmetry, the improper operation $S_4$ to $\bm{D}^{(3)}$ is represented by $R^{(3)}[S_4]=i\tau_2$, which leads to 
$\tau_2 \hat{\bm{\chi}}^{(3)} \tau_2 = \hat{\bm{\chi}}^{(3)}$.
$C_{2x}$ transforms $\bm{D}^{(3)}$ according to $R^{(3)}[C_{2x}]=\tau_3$, hence, 
$\tau_3 \hat{\bm{\chi}}^{(3)}
\tau_3= \hat{\bm{\chi}}^{(3)}$. 
Therefore, the $D_{2d}$ symmetry constraints
$\bm{\hat{\chi}}^{(3)}=\chi_0^{(3)} I$
and $\Delta J^{(3)}_c(\phi)$  as
\begin{eqnarray}
\Delta J^{(3)}_c(\phi)=\frac{2}{\pi} B\chi_0^{(3)}\cos (3\phi-\phi_B).
\label{eq:d2d_oct}
\end{eqnarray}

\textit{Microscopic calculations.--} 
To validate the above symmetry analysis, we study the superconducting diode effect in a 2D square lattice with Rashba ($C_{4v}$)
and Dresselhaus ($D_{2d}$) SO couplings, both of which break the inversion symmetry. 
The analysis below is based on a mean-field treatment of the following Hamiltonian, which consists of the kinetic energy, the magnetic field,
the SO coupling \cite{Dresselhaus1955,Manchon2015a,Wu2004,Zhang2025,Bychkov1984,Bonesteel1992}, and the interaction term. 
The kinetic energy part reads, 
$H_0=\sum_{\bm{k}\sigma}\xi_{\bm{k}\sigma}c_{\bm{k}\sigma}^{\dagger}c_{\bm{k}\sigma}$, in which the band dispersion
$\xi_{\bm{k}\sigma}=-2t\left[\cos(k_x)+\cos(k_y)\right]-\mu$ with the hopping amplitude $t = 1$ and chemical potential $\mu = -1$.
The magnetic field term $H_{\rm B}=-\sum_{\bm{k}\sigma\sigma'}\bm{B}\cdot \bm{\tau}_{\sigma,\sigma'}c_{\bm{k}\sigma}^{\dagger}c_{\bm{k}\sigma'}$, where $\bm{B} = (B_x, B_y)$ denotes the applied in-plane magnetic field. 
The SO coupling term is
\begin{eqnarray}
H_{\rm SOC}=\sum_{\bm{k}\sigma\sigma'}\bm{g}(\bm{k})\cdot \bm{\tau}_{\sigma\sigma'}c_{\bm{k}\sigma}^{\dagger}c_{\bm{k}\sigma'}
\label{eq:SO}.
\end{eqnarray}
The interaction term reads
$H_{\rm int}=\sum_{\bm{k}\bm{k}'}V_{\bm{k}\bm{k}'}c_{\frac{\bm{q}}{2}+\bm{k}\uparrow}^{\dagger}c_{\frac{\bm{q}}{2}-\bm{k}\downarrow}^{\dagger}c_{\frac{\bm{q}}{2}-\bm{k}'\downarrow}c_{\frac{\bm{q}}{2}+\bm{k}'\uparrow}$, where for simplicity the pairing potential is taken to be momentum independent with the form $V_{\bm{k}\bm{k}'}=-V$, where $V = 1.5$.

Consider the mean-field Hamiltonian for the current carrying superconducting state, 
\begin{eqnarray}
	H_{\rm MF}(\Delta_{\bm{k}}^{\bm{q}})=\frac{1}{2}\sum_{\bm{k}}\psi^{\dagger}(\bm{k},\bm{q})H(\bm{k},\bm{q})\psi(\bm{k},\bm{q}),
    \label{eq:meanHam}
\end{eqnarray}
where $\bm{q}$ is the center of mass momentum of pairing and the Nambu spinor is defined by
$\psi(\bm{k},\bm{q})=\left(c_{\bm{q}/2+\bm{k}\uparrow}^{\dagger}~c_{\bm{q}/2+\bm{k}\downarrow}^{\dagger}~c_{\bm{q}/2-\bm{k}\uparrow}~c_{\bm{q}/2-\bm{k}\downarrow}\right)^{\dagger}.
$
Here, $c_{\bm{k}\sigma}$ denotes the annihilation operator for an electron with momentum $\bm{k}$ and spin component $\sigma$. 
The Bogoliubov–de Gennes (BdG) Hamiltonian matrix $H(\bm{k}, \bm{q})$ takes the form of
\begin{eqnarray}
	H(\bm{k},\bm{q})=\left(\begin{array}{cc}
		H_{N}(\bm{q}/2+\bm{k})&\Delta_{\bm{k}}^{\bm{q}}(i\tau_2)\\
		(\Delta_{\bm{k}}^{\bm{q}})^*(-i\tau_2)&-H_{N}^{T}(\bm{q}/2-\bm{k})
	\end{array}\right),
    \nonumber
\end{eqnarray}
where $H_N(\bm{k})$ describes the non-interacting normal-state 
Hamiltonian including the dispersion, the magnetic field and SO coupling terms, and $\Delta_{\bm{k}}^{\bm{q}}$ is the superconducting gap function.
The superconducting gap equation can be expressed as
\begin{eqnarray}
	\Delta_{\bm{k}}^{\bm{q}}=\sum_{\bm{k}'}V_{\bm{k}\bm{k}'}\sum_{j=1-4}f(E_{j\bm{k}'})V_{4j}^*V_{1j},
\end{eqnarray}
where the matrix $V$ diagonalizes the BdG Hamiltonian via  $V^{\dagger}H(\bm{k},\bm{q})V=\Lambda_{\bm{k}}$ with $\Lambda_{j\bm{k}} = E_{j\bm{k}}$
representing the quasi-particle energies. 
The free energy difference $F(\bm{q})$ between 
the normal and superconducting states can be calculated in the zero temperature limit.
The supercurrent $\bm{j}(\Delta_{\bm{k}}^{\bm{q}})$
follows from $\bm{j}(\Delta_{\bm{k}}^{\bm{q}})=2\partial_{\bm{q}}F(\bm{q})$ \cite{Daido2022a,Yuan2022a}, where the coefficient 2 is due to the fact that $\bm{q}$ is the momentum of Cooper pairs. 
The critical current density $J_c^{ } (\phi)$ is taken as the maximum of $\bm{j}$ when varying the magnitude of $\bm{q}$ along the direction of $\phi$.

\textit{Rashba spin-orbit coupling.--} 
We assign the Rashba SO coupling as $\bm{g}(\bm{k})=\alpha_g (-\sin k_y,\sin k_x,0)$ in Eq. (\ref{eq:SO}), which breaks inversion $\mathcal{P}$ but maintains $C_{4z}$. 
When both the Rashba SO coupling and the magnetic field $\bm{B}$ are present, both $\mathcal{P}$ and $\mathcal{T}$ are broken, nevertheless, the superconducting diode effect could still be forbidden by other symmetries. 
If $\bm{B}\parallel \hat z$, then the 2-fold rotation symmetry around the $z$-axis $C_{2z}$ ensures the absence of the diode effect, {\it i.e.}, $\Delta J_c(\phi)=0$.
Hence, below we consider the situation
of in-plane magnetic field.

When the in-plane $\bm{B}$ is parallel to the current, a mirror reflection across a vertical plane perpendicular to the current reverses the current while leaving $\bm{B}$ invariant. When 
$\bm{B}$ is perpendicular to the current, no symmetry operation simultaneously reverses the current and preserves $\bm{B}$, thus allowing the superconducting diode effect.

\begin{figure}[t]
\begin{center}
\includegraphics[width=8cm]{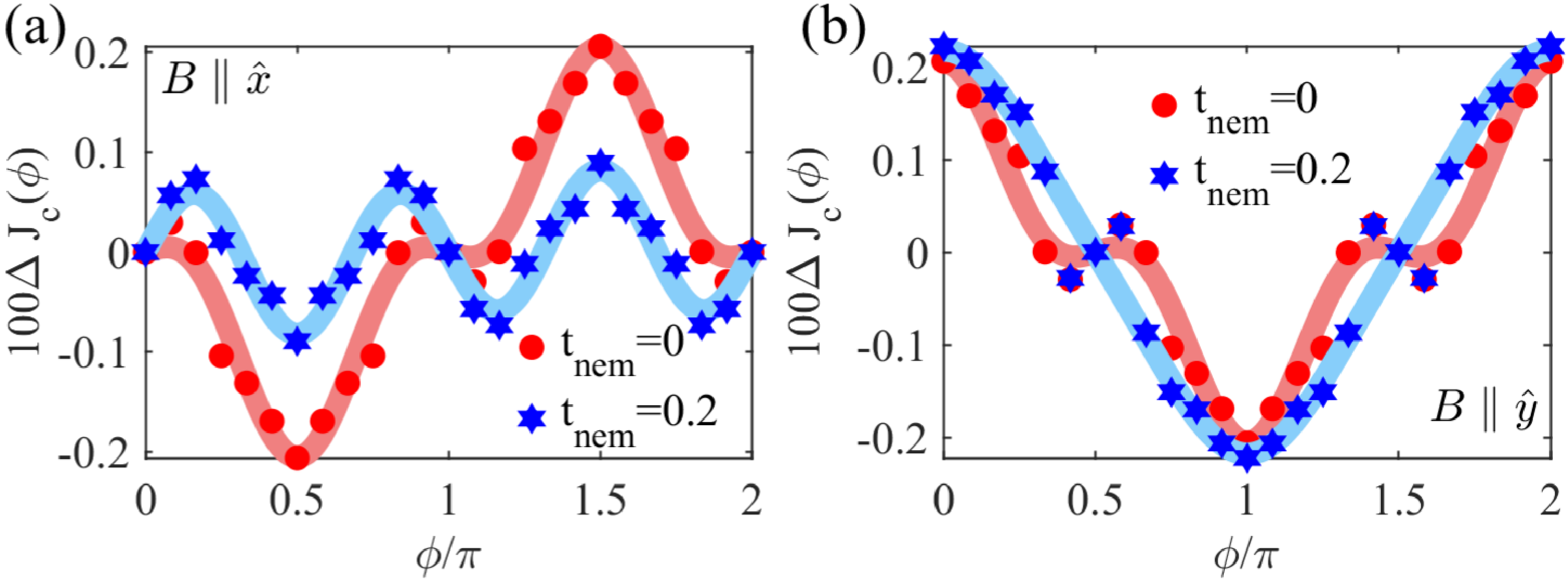}
\end{center}
\caption{The non-reciprocal critical current $\Delta J_c(\phi)=J_c(\phi)-J_c(\phi+\pi)$. 
The in-plane magnetic field $\bm{B}$ is along the $x$- and $y$-axis in (a) and (b), respectively, and its magnitude 
$B=0.03$. 
Rashba SO coupling is set as 
$\alpha_g = 0.3$.
$J_c(\phi)$ is the critical current calculated 
as the maximum supercurrent obtained by varying the pairing momentum $\bm{q}$ in Eq. (\ref{eq:meanHam}).
In (a) and (b), the $\Delta J_c(\phi)$ data
are fitted by the Fourier components at the 1st and 3rd order harmonics. 
The red dots represent the case of the tetragonal symmetry manifested as a relative shift of $\pi/2$ in (a) and (b);
the blue dots represent the case with orthorhombic distortion.
}
\label{fig1}
\end{figure}

To further substantiate this point, we use mean-field theory to calculate the critical current for both the tetragonal and orthorhombic (nematic) cases. The nematic distortion breaks $C_{4z}$
symmetry and is modeled by a nematic hopping term as 
\begin{eqnarray}
H_{\rm nem}=-2t_{\rm nem}\sum_{\bm{k}\sigma}
f(\bm{k})c_{\bm{k}\sigma}^{\dagger}c_{\bm{k}\sigma},
\label{eq:nematic}
\end{eqnarray}
with $f(\bm{k}) =\cos k_x-\cos k_y$ \cite{Lederer2015a}.
In this case, the response tensors 
$\hat{\bm{\chi}}^{(1)}$ and 
$\hat{\bm{\chi}}^{(3)}$ are no longer
represented by Eqs. (\ref{eq:c4v_dip})
and (\ref{eq:d2d_dip}), respectively.

In Fig. \ref{fig1} (a), the applied magnetic field $\bm{B}\parallel \hat x$ and the current flows along the direction of the azimuthal angle $\phi$. 
By definition, $\Delta J_c(\phi)=-\Delta J_c(\pi+\phi)$.
Furthermore, the reflection symmetry $M_{yz}$ maintains the magnetic field $\bm{B}$ but changes the direction of current from $\phi$ to 
$\pi-\phi$, yielding $\Delta J_c(\phi)=\Delta J_c(\pi -\phi)$.
Hence, the polarizability is odd under 
$\phi \to \pi+\phi$ and even under $\phi\to \pi-\phi$.
Based on these properties, $\Delta J_c(\phi)$ is expanded as 
\begin{eqnarray}
\Delta J_c(\phi)=\sum_{k=1}^{\infty}(-)^{k}s_{2k-1}\sin(2k-1)\phi.
\end{eqnarray}
Hence, the diode effect is absent at $\phi=0$ and $\pi$.
$\Delta J_c(\phi)$ in Fig. \ref{fig1} (a) can be well fitted by the leading two components of $-s_1\sin \phi+s_3\sin 3\phi$ both in the absence and presence of nematicity. 
Based on the definitions in Eq. (\ref{eq:D1}), we extract $\bm{D}^{(1)}=(0, -\frac{\pi}{2}s_1)$ and $\bm{D}^{(3)}=(0, \frac{\pi}{2}s_3)$, yielding the response tensor components 
$\chi_{11}^{(1)}=0,~\chi_{21}^{(1)}=-s_1 \frac{\pi}{2B}$, and
$\chi_{11}^{(3)}=0,~\chi_{21}^{(3)}= s_3 \frac{\pi}{2B}$.
For $\bm{B}\parallel \hat y$ in Fig. \ref{fig1} (b), a similar symmetry analysis gives the expansion
$\Delta J_c(\phi)=\sum_{k=1}^{\infty}c_{2k-1}\cos(2k-1)\phi$. In the absence of nematicity, a $\frac{\pi}{2}$ rotation, to the results of $\bm{B}\parallel \hat x$, gives $c_k=s_k$.
Nevertheless, the nematicity term breaks the 
$C_{4z}$ rotation symmetry, $c_k$ and $s_k$ become independent. 
In this case, we extract that $\bm{D}^{(1)}=(c_1 \frac{\pi}{2},0)$ and 
$\bm{D}^{(3)}=(c_3 \frac{\pi}{2},0)$,
yielding $\chi_{22}^{(1)}=0,~\chi_{12}^{(1)}=c_1 \frac{\pi}{2B}$, and 
$\chi_{22}^{(3)}=0,~\chi_{12}^{(3)}=c_3 \frac{\pi}{2B}$.

Then the response tensor  takes the form of
$\hat{\bm{\chi}}^{(1)}=\chi_{nm} \tau_1+\chi_{C_4} i\tau_2,~\hat{\bm{\chi}}^{(3)}=\chi_{nm} i\tau_2+\chi_{C_4} \tau_1$ with $\chi_{nm}=\frac{\pi}{4B}(c_k- s_k)$
and $\chi_{C_4}=\frac{\pi}{4B}(s_k+c_k)$.
Hence, $\hat{\bm{\chi}}^{(1)}$ is fully anti-symmetric in the presence of the tetragonal symmetry. 
On the other hand, nematicity induces a finite symmetric component $\chi_{nm}\tau_1$, reflecting the broken rotational symmetry. The response tensor thus serves as a probe for nematicity in real materials. Additionally,  the nematicity-induced enhancement or suppression of the extrema in $\Delta J_c$, as shown in Fig. \ref{fig1}, suggests a possible route to tune the superconducting diode efficiency by varying the nematicity strength.

\begin{figure}[t]
\begin{center}
\includegraphics[width=8cm]{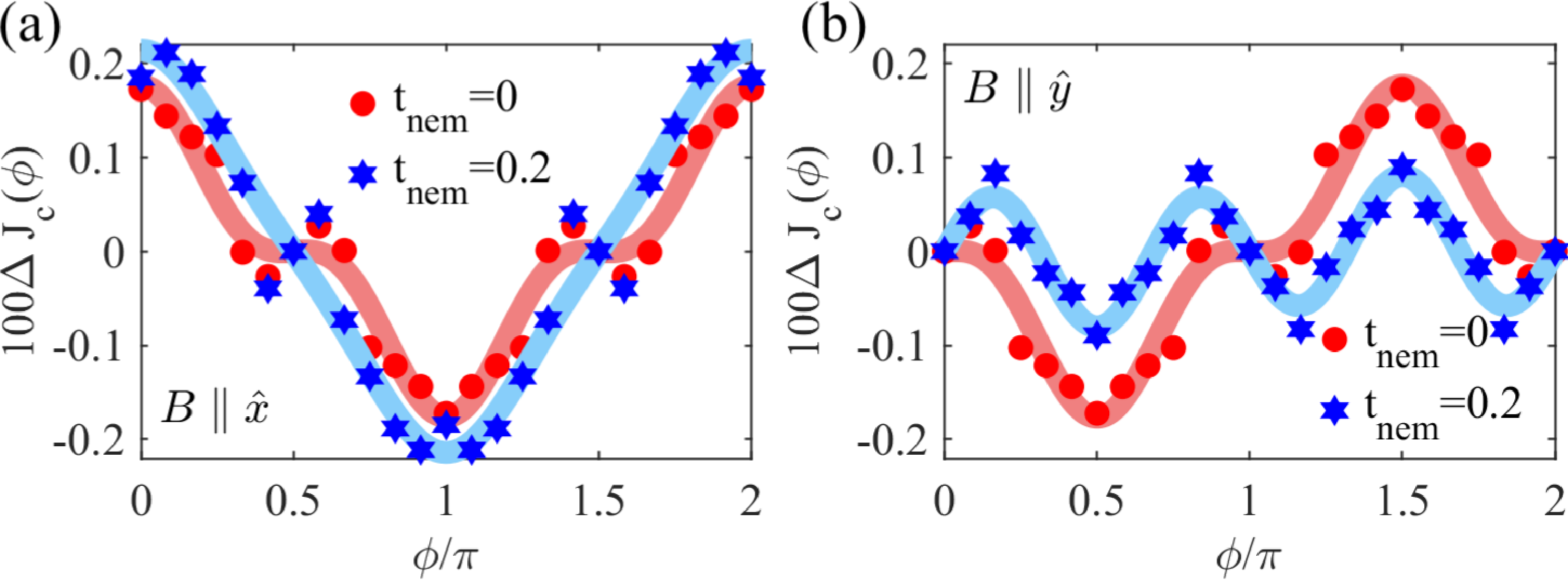}
\end{center}
\caption{The parallel calculations for the non-reciprocal critical current $\Delta J_c(\phi)$ in the presence of Dresselhaus SO coupling.
The Dresselhaus SO coupling is set as $\alpha_g = 0.3$.
Other parameter settings in (a) and (b) are the same as in Fig. \ref{fig1} (a) and (b), respectively.
In (a) and (b), the red dots represent the case of
$S_4$ symmetry,
and the blue dots represent the case with orthorhombic distortion.
	}
	\label{fig2}
\end{figure}

\textit{Dresselhaus spin-orbit coupling.--} 
We similarly analyze the system with the Dresselhaus SO coupling, characterized by $\bm{g}(\bm{k})=\alpha_g(\sin k_x,-\sin k_y,0)$ in Eq. (\ref{eq:SO}) and other terms of Hamiltonian are the same as that in the Rashba case.  
When the system remains tetragonal, it possesses the $D_{2d}$ symmetry, which breaks inversion $\mathcal{P}$, 
while preserving the twofold rotational symmetries $C_{2z,2x,2y}$.
Compared to the Rashba case, the role of 
$C_{4z}$ is replaced by $S_4$.

The Dresselhaus SO coupling results in different symmetry patterns of superconducting diode effect. 
For $\bm{B}\parallel \hat z$, $C_{2z}$ symmetry reverses the in-plane current, thereby forbidding the superconducting diode effect.
Similarly, if $\bm{B}$ is along the $x$, or $y$-axis, $\bm{J}$ flows perpendicularly to $\bm{B}$, then the $C_{2x}$ or $C_{2y}$ symmetry operation reverses the current, again preventing the diode effect.
However, when the magnetic field is parallel to the current, say, 
$\bm{B} \parallel \hat{x} \parallel \bm{I}$, there does not exist a symmetry to reverse the current but maintain
$\bm{B}$, leaving the diode effect symmetry-allowed. Thus,
in contrast to the Rashba case, the Dresselhaus-induced superconducting diode effect emerges only when $\bm{B}$ has a component parallel to $\bm{J}$.


We perform self-consistent mean-field calculations of 
the current density $\bm{j}(\bm{q})$ for both the $S_4$  and orthorhombic (nematic) case.
For the case $\bm{B} \parallel \hat x$, the $C_{2x}$ symmetry maintains $\bm{B}$ invariant but changes 
the azimuthal angle $\phi$ of $\bm{J}$ to $-\phi$, which constrains the expansion of
$\Delta J_c$ to
\begin{eqnarray}
\Delta J_c(\phi)=\sum_{k=1}^{\infty} c^\prime_{2k-1}\cos(2k-1)\phi.
\end{eqnarray}
This relation confirms the absence of the diode effect as $\bm{B} \perp \bm{I}$. As shown in Fig.~\ref{fig2} (a), the angular dependence $\Delta J_c(\phi)$ can be accurately fitted by first two Fourier components $\Delta J_c(\phi) = c_1^\prime \cos\phi + c_3^\prime \cos 3\phi$. Consequently, $\bm{D}^{(1)} = \left(c_1^\prime \frac{ \pi}{2}, 0 \right)$ and $\bm{D}^{(3)} = \left(c_3^\prime \frac{ \pi}{2}, 0 \right)$, yielding the response tensor components $\chi^{(1)}_{21} = 0,~\chi_{11}^{(1)} = \frac{c_1^\prime \pi}{2B}
$ and $\chi^{(3)}_{21} = 0,~\chi_{11}^{(3)} = \frac{c_3^\prime \pi}{2B}$.
Similarly, for $\bm{B}\parallel \hat y$, we arrive at $\Delta J_c(\phi)=\sum_{k=1}^{\infty} (-)^{k}s^\prime_{2k-1}\sin(2k-1)\phi$.
The results shown in Fig.~\ref{fig2}(b) can be fitted as $\Delta J_c(\phi) = -s^\prime_1 \sin\phi + s^\prime_3 \sin 3\phi$, leading to $\chi_{12}^{(1)} = 0,~\chi_{22}^{(1)} = -\frac{s_1^{\prime}\pi}{2B}$ and $\chi_{12}^{(3)} = 0,~\chi_{22}^{(3)} =\frac{s_3^{\prime}\pi}{2B}$. 
In the presence of the $S_4$ symmetry, $s^\prime_k=c^\prime_k$, giving rise to a diagonal response tensor: 
$\hat{\bm{\chi}}^{(1)}=\frac{s_1^{\prime}\pi}{2B}\tau_3,~\hat{\bm{\chi}}^{(3)}=\frac{s_3^{\prime}\pi}{2B}I$. 
Nevertheless, when nematicity is introduced ($t_{\mathrm{nem}} \neq 0$), $c^\prime_k\neq s^\prime_k$.
The dipole and octupolar response tensor can then be decomposed as
$\bm{\hat \chi}^{(1)}=\chi_{nm} I +\chi_{S_4} \tau_3,~\bm{\hat \chi}^{(3)}=\chi_{nm}\tau_3 +\chi_{S_4} I$
with $\chi_{nm}=\frac{\pi}{4B}(c_k^{\prime}-s_k^{\prime})$ and $\chi_{S_4}=\frac{\pi}{4B}(c_k^{\prime}+s_k^{\prime})$.
Here, $\chi_{nm}$ serves as a direct measure of nematicity in the superconducting state.

%

\begin{figure}[t]
\begin{center}
\includegraphics[width=8cm]{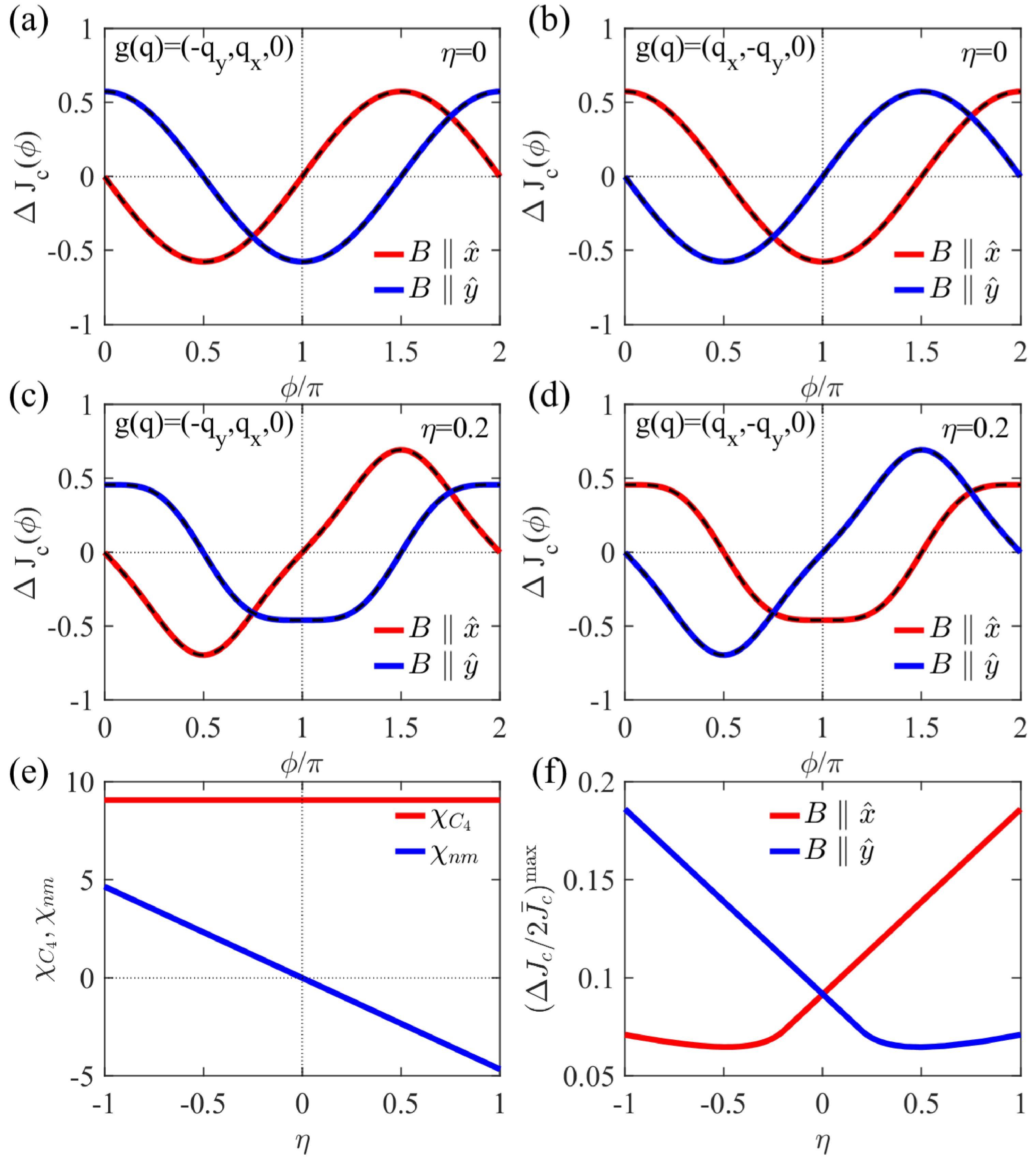}
\end{center}
\caption{The angular dependences of the non-reciprocal relations of critical currents $\Delta J_c(\phi)$.
(a) and (c) are for the cases of Rashba SO coupling with the nematic parameter $\eta=0$ and $0.2$, respectively. 
(b) and (d) are for the Dresselhaus SO coupling with the nematic parameter $\eta=0$ and $0.2$, respectively. 
The red and blue lines are for the in-plane magnetic-field $\bm{B}$ along the $x$- and $y$-axis, respectively.
(e) The $\chi_{C4}$ component and $\chi_{nm}$ 
component with the nematicity strength $\eta$ for the Rashba case.
(f) The maximum of superconducting diode 
efficiency $\Delta J_c/2\Bar{J}_c$ for the Rashba cases, where $\Bar{J}_c=\frac{1}{2}[J_c(\phi)+J_c(\phi+\pi)]$, as functions of the nematic strength $\eta$, for different magnetic field directions, respectively. We also note that the Dresselhaus case exhibits almost identical features to those in (e) and (f).
	}
	\label{fig3}
\end{figure}

The $\chi_{nm}$ components signal the breaking of $C_{4z}$ and $S_4$ symmetries for Rashba and Dresselhaus SO couplings, respectively. Their emergence is therefore expected to be robust against microscopic details of the nematic order, including its form factor ($d_{x^2-y^2}$ or $d_{xy}$), its origin in the particle-hole or particle-particle channel, and whether it is spin-dependent or spin-independent.

\textit{Ginzburg--Landau analysis.--}
Additionally, the non-reciprocal relation of critical currents $\Delta J_c(\phi)$ can also be derived by using Ginzburg–Landau theory, with the free energy density expressed as $f(\bm{q},\Delta)=\alpha_{\bm{q}}|\Delta|^2+\frac{1}{2}\beta |\Delta|^4$, where $\alpha_{\bm{q}}=t+a_0q^2-(b_0-(b_{1x}q_x^2+b_{1y}q_y^2))\bm{g}(\bm{q})\cdot \bm{B}$ \cite{Yuan2022a,Daido2022b}. 
The Rashba SO coupling takes the form  $\bm{g}(\bm{q})=(-q_y,q_x,0)$, while the Dresselhaus SO coupling is represented by $\bm{g}(\bm{q})=(q_x,-q_y,0)$. 
In Figs.~\ref{fig3}(a), (b), (c) and (d),  the computed $\Delta J_c(\phi)$ is shown for various orientations of the magnetic field for $\eta=0$ and $0.2$, where $\eta=\frac{b_{1y}-b_{1x}}{b_{1y}+b_{1x}}$ denotes the strength of the nematicity in the free energy. 
These results yield a similar form for $\bm{\hat{\chi}}$ as that obtained from the mean-field calculations. 
We also note that the $\chi_{C4}$
response for the Rashba SO coupling and $\chi_{S_4}$ one for the Desselhaus SO coupling remains insensitive to the nematic strength $\eta$, while the 
symmetry breaking indicator 
$\chi_{nm}$ exhibits a linear dependence, as shown in Fig.~~\ref{fig3}(e). 
This behavior leads to either an enhancement or suppression of the maximum for the superconducting diode efficiency $\Delta J_c/ 2\Bar{J}_c$ in Fig.~~\ref{fig3}(f), where $\Bar{J}_c=\frac{1}{2}(J_c(\phi)+J_c(\phi+\pi))$.
Consequently, nematicity serves as an effective tuning parameter for controlling the superconducting diode efficiency.

%

\textit{Conclusion and Discussion.--} 
In summary, we establish a unified framework for characterizing the superconducting diode effect
through  response tensors $\hat{\bm{\chi}}^{(n)}$, which relate the multipole vector $\bm{D}^{(n)}$ encoding the angular dependence of the critical current to an applied in-plane magnetic field $\bm{B}$. For systems with Rashba SO couplings and $C_{nv}$ symmetry ($n=3,4,6$), the linear-response tensor $\hat{\bm{\chi}}^{(1)}$ exhibits an antisymmetric structure analogous to that of the Hall conductivity in the normal state.
Furthermore, deviations from antisymmetry in $\hat{\bm{\chi}}^{(1)}$ provide a sensitive probe of superconducting nematicity, although they do not reveal the microscopic origin of the nematic order. For systems with the $D_{2d}$ exhibiting the Dresselhaus SO coupling, $\hat{\bm{\chi}}^{(1)}$ exhibits a diagonal traceless form, and the nematicity brings the trace part to the response. 
Additionally, the $C_{4v}$ and $D_{2d}$ symmetries give rise to symmetric off-diagonal and identity forms of $\hat{\bm{\chi}}^{(3)}$, respectively.
We therefore propose that $\hat{\bm{\chi}}^{(n)}$ should be experimentally measurable in various systems exhibiting the superconducting diode effect. 
Its structure may reflect intrinsic properties of the underlying system, such as nematic order, and 
the specific form of SO coupling. 


We can also rotate the in-plane magnetic field, then the harmonic analysis with respect to both $\phi$ and the angle $\phi_B$ can yield the information of non-linear responses to dipole and octupole components. 
The related discussion about symmetry constraint of non-linear response is presented at the End Matter. 
Thus, our framework provides an experimentally accessible route to identifying dipolar and octupolar contributions at both linear and nonlinear orders in superconducting diode measurements.

{\it Acknowledgments.--} 
C.W. was supported by the National Natural Science Foundation of China
under the Grant Nos. 12234016 and 12550402, 12574274.
J.W. acknowledges National Key R\&D Program of China (No. 2024YFA1408102 and No. 2023YFA1406400), Research Center for Industries of the Future (RCIF project No. WU2023C001 ) at Westlake University, the National Natural Science Foundation of China (Grant No. 12174318), the Zhejiang Provincial Natural Science Foundation of China (Grant No. XHD23A2002). 
This project is also supported by
the New Corner stone Science Foundation.
The authors also acknowledge the computation resource provided by Westlake HPC Center.

\bibliography{refs.bib}

\onecolumngrid
\clearpage
\appendix*
\begin{center}
	{\bf\large End Matter}
\end{center}
\twocolumngrid
\setcounter{equation}{0}
We consider the non-linear response tensors to the $\bm{D}^{(1)}$ and $\bm {D}^{(3)}$ components to the in-plane magnetic fields.

\textit{Response tensors for 3-fold symmetries.--}
Consider a 2D lattice system with the $C_{3v}$ symmetry. 
Since the angular momentum is conserved modulo 3, the non-linear response at the 2nd-order of in-plane components of $B$ can also generate  $\bm{D}^{(1)}$.
However, if the superconductivity is time-reversal invariant by itself, this response is forbidden by time-reversal symmetry. 
Nevertheless, if the system itself is time-reversal symmetry breaking, either due to chiral superconductivity or by proximity to magnetism, we can write down 
\begin{eqnarray}
   D_i^{(1)} = \sum_{jk} \chi^{(1)}_{i,jk}B_j B_k m_z,
\end{eqnarray}
where $m_z$ represents a magnetic moment-like
time-reversal symmetry breaking property possessed by the system, whose orentation is along $z$-axis. 

As for a point-group operation $O$ of $C_{3v}$,
the transformation to $\chi^{(1)}_{i,jk}$ reads
\begin{eqnarray}
\chi^{(1)}_{i,jk}&=&{\rm sign}[O]\sum_{i'j'k'}R_{i'i}^{(1)}[O]~\chi_{i',j'k'}
R^{(1)}
[O]_{j'j} R^{(1)}_{k'k}[O].
\nonumber 
\label{eq:rank_3}
\end{eqnarray}
The mirror symmetry $M_{xz}$ is represented by $R^{(1)}(M_{xz})=\tau_3$,
$\chi_{i,jk}^{(1)} = -\sum_{i'j'k'} (\tau_3)_{i'i}\, \chi^{(1)}_{i',j'k'}(\tau_3)_{j'j}(\tau_3)_{k'k}$,
yielding
$\chi^{(1)}_{1,xx} = \chi^{(1)}_{1,yy} = \chi^{(1)}_{2,xy}=0.$
Then Eq. (\ref{eq:rank_3}) is reduced to 
\begin{eqnarray}
D^{(1)}_1 &=& 2\chi^{(1)}_{1,xy} B_x B_y m_z,    \nonumber \\
D^{(1)}_2 &=& (\chi^{(1)}_{2,xx} B_x^2 + 
\chi^{(1)}_{2,yy}  B_y^2) m_z.
\end{eqnarray}
The 3-fold rotation $C_{3z}$
represented by $R^{(1)}[C_{3z}] =-\frac{1}{2}I +
\frac{\sqrt 3}{2} (-i\tau_2)$
 , which leads to $\chi_{1,xy}^{(1)}=
\chi_{2,xx}^{(1)}=-\chi_{2,yy}^{(1)}=\chi^{(1)}_{B^2}$.
Under complex combinations, we organize the 2nd order response as
\begin{eqnarray}
D^{(1)}_1+iD^{(1)}_2
=i\chi^{(1)}_{B^2}(B_x-iB_y)^2 m_z.
\end{eqnarray}
This response results in the first harmonic distribution of $\Delta J_c^{(1)}(\phi)$ as
\begin{eqnarray}
\Delta J_c^{(1)}(\phi)
= \frac{2}{\pi}
B^2 \chi_{B^2}^{(1)} 
\sin(\phi +2\phi_B)m_z.
\label{eq:2nd_harm}
\end{eqnarray}

The rotationally invariant $m_z$ by itself can generate the octupole component $\bm D^{(3)}$ as
$D_i^{(3)}=\chi^{(3)}_i m_z$.
The mirror symmetry $M_{xz}$ is represented by
$R^{(3)}[M_{xz}]=R^{(1)}[M_{xz}]=\tau_3$,
we arrive at $\chi_1=0$.
Hence, 
\begin{eqnarray}
D^{(3)}_1=0, \ \ \,
D^{(3)}_2=\chi^{(3)}_2 m_z,
\end{eqnarray}
which leads to the 3rd harmonic angular distribution as
\begin{eqnarray}
\Delta J_c^{(3)}(\phi)
=\frac{2}{\pi}  \chi^{(3)}_2\sin 3\phi ~m_z.
\label{eq:3rd_harm1}
\end{eqnarray}


\textit{Cubic response tensors to 
the dipole and octupole components.--}
Below we will consider the non-linear response tensors at the cubic order under 6-fold and 4-fold symmetries. 
We summarize their common properties
due to $M_{xz}$ below.
The rank-4 tensor $\chi_{i,jkl}$ 
for octupole $(n=3)$ and dipole 
$(n=1)$ components are expressed in a unified way:
\begin{eqnarray}
   D_i^{(n)} = \sum_{jkl} \chi_{i,jkl}^{(n)} B_j B_k B_l ,
\label{eq:cubic_resp}
\end{eqnarray}
where $i=1,2$ and $j,k,l$ take the values of $x,y$.
As for a point-group operation $O$,
the transformation reads
\begin{eqnarray}
\chi_{i,jkl}^{(n)} &=&{\rm sign}[O]\sum_{i'j'k'l'}R_{i'i}^{(n)}[O]~\chi_{i',j'k'l'}^{(n)}
\nonumber \\
&\times& 
~R^{(1)}_{j'j}
[O] R^{(1)}_{k'k}[O]  R^{(1)}_{l'l}[O].
\label{eq:rank_4}
\end{eqnarray}
The mirror symmetry $M_{xz}$ is represented by
$R^{(3)}[M_{xz}]=R^{(1)}[M_{xz}]=\tau_3$,
yielding
$
\chi_{1,xxx}^{(n)} = \chi_{2,yyy}^{(n)} = \chi_{1,xyy}^{(n)} = \chi_{2,xxy}^{(n)} = 0.
$
Then Eq. (\ref{eq:cubic_resp}) is reduced to 
\begin{eqnarray}
D^{(n)}_1 &=& 3\chi^{(n)}_{1,xxy} B_x^2B_y + 
\chi_{1,yyy}^{(n)} B_y^3,    \nonumber \\
D^{(n)}_2 &=& \chi_{2,xxx}^{(n)} B_x^3 + 3\chi_{2,xyy}^{(n)} B_x B_y^2.
\label{eq:cubic_Mzx}
\end{eqnarray}


\textit{Response tensors for 6-fold symmetries.--}
Next consider the cubic order response to $\bm{D}^{(3)}$ under the $C_{6v}$ symmetry. 
$C_{6z}$ is represented by 
$R^{(1)}[C_{6z}] =\frac{1}{2}I +
\frac{\sqrt 3}{2} (-i\tau_2)$, 
and $R^{(3)}[C_{6z}]=-I$, 
which leads to 
$\chi_{1,xxy}^{(3)}
=-\chi_{1,yyy}$, 
$\chi_{2,xxx}^{(3)}=-\chi_{2,xyy}^{(3)}$ in Eq. (\ref{eq:cubic_Mzx})
by setting $n=3$. 
Thus we arrive at the
harmonics of the critical current distribution as 
\begin{eqnarray}
    \Delta J_c^{(3)}(\phi)&=&\frac{2}{\pi} B^3 (\chi_{2,xxx}^{(3)} \sin 3\phi \cos 3\phi_B
    \nonumber \\
    &&~~~~~-\chi_{1,yyy}^{(3)} \cos 3\phi \sin 3\phi_B
    ).
\label{eq:3rd_harm2}
\end{eqnarray}
When the current is along the $x$ direction, we have $\Delta J_c^{(3)}(0)=-2/\pi B^3\chi_{1,yyy}^{(3)}\sin 3\phi_B$. 
Under complex combinations, we organize the cubic order response 
as
\begin{eqnarray}
D^{(3)}_1+iD^{(3)}_2
&=&i\chi^{(3)} (B_x+iB_y)^3 \nonumber \\
&+& 
i\tilde{\chi}^{(3)}(B_x-iB_y)^3,
\end{eqnarray}
where $\chi^{(3)}=\frac{1}{2}(\chi_{1,yyy}^{(3)}+\chi_{2,xxx}^{(3)})$,
 $\tilde{\chi}^{(3)}=\frac{1}{2}(-\chi_{1,yyy}^{(3)}+\chi_{2,xxx}^{(3)})$.

\textit{Cubic response tensors for $C_{4v}$.--}
The cubic order response in this case contributes to both the octupole and dipole components since angular momentum is conserved modulo 4.
The analysis can be performed in parallel to the case of $C_{6v}$.
Again the mirror symmetry $M_{xz}$ 
yields Eq. (\ref{eq:cubic_Mzx})
with $n=3$.
$C_{4z}$ represented by $R^{(3)}[C_{4z}]=-R^{(1)}[C_{4z}]=i\tau_2$, yields
$~\chi_{1,yyy}^{(3)} = \chi_{2,xxx}^{(3)},
~\chi_{2,xyy}^{(3)} = \chi_{1,xxy}^{(3)}$.
Under complex combinations, we organize the cubic response as
\begin{eqnarray}
D^{(3)}_1+iD^{(3)}_2
&=& i\chi^{(3)} (B_x+iB_y)^3 
\nonumber \\
&+&i \tilde{\chi}^{(3)} B^2
(B_x-i B_y),
\end{eqnarray}
where $\chi^{(3)}=\frac{1}{4}(\chi_{2,xxx}^{(3)}-3\chi_{1,xxy}^{(3)})$,
 $\tilde{\chi}^{(3)}=\frac{3}{4}(\chi_{2,xxx}^{(3)}+\chi_{1,xxy}^{(3)})$.
We obtain the harmonics of the critical  current distribution as
\begin{eqnarray}
 \Delta J^{(3)}_c(\phi)&=& 
 \frac{2}{\pi}B^3
[\chi^{(3)}\sin(3\phi-3\phi_B)\nonumber \\
&&~~~~~~~+ \tilde {\chi}^{(3)} \sin (3\phi+\phi_B) 
].
\label{eq:3rd_harm3}
\end{eqnarray}


Similarly, the cubic response also contributes to the dipole component,
yielding Eq. (\ref{eq:cubic_Mzx})
with $n=1$.
Furthermore, the $C_{4z}$ symmetry gives rise to $~\chi_{1,yyy}^{(1)} = -\chi_{2,xxx}^{(1)},
~\chi_{2,yyx}^{(1)} = -\chi_{1,xxy}^{(1)}$, such that
the contribution can be organized under complex combinations as
\begin{eqnarray}
D^{(1)}_1+iD^{(1)}_2
&=& i \chi^{(1)} B^2
(B_x+i B_y)
\nonumber \\
&+& i\tilde{\chi}^{(1)} (B_x-iB_y)^3, 
\end{eqnarray}
where $\chi^{(1)}=\frac{3}{4}(-\chi_{1,xxy}^{(1)}+\chi_{2,xxx}^{(1})$,
 $\tilde{\chi}^{(1)}=\frac{1}{4}(3\chi_{1,xxy}^{(1)} +\chi_{2,xxx}^{(1)})$. 
Thus, we have
\begin{eqnarray}
  \Delta J_c^{(1)} (\phi)&=&\frac{2}{\pi}B^3[\chi^{(1)}\sin(\phi-\phi_B)\nonumber\\
&&~~~~~~+\tilde{\chi}^{(1)}\sin(\phi+3\phi_B) ].
\end{eqnarray}


\textit{Cubic response tensors for $D_{2d}$.--}
Again we consider the cubic response to both octupole ($n=3$) and dipole ($n=1$)
components.
The $C_{2x}$ symmetry yields
$\chi_{1,yyy}^{(n)} = \chi_{2,xxx}^{(n)} = \chi_{1,xxy}^{(n)} = \chi_{2,xyy}^{(n)} = 0$.
Then Eq. (\ref{eq:cubic_resp}) 
becomes
\begin{eqnarray}
D^{(n)}_1 &=& \chi^{(n)}_{1,xxx} B_x^3 + 
3 \chi_{1,xyy}^{(n)} B_x B_y^2,    \nonumber \\
D^{(n)}_2 &=& 3\chi_{2,xxy}^{(n)} B_x^2 B_y + \chi_{2,yyy}^{(n)} 
B_y^3.
\end{eqnarray}
The $S_4$ symmetry operation, represented by $R^{(1)}[S_4]=-R^{(3)}[S_4]=-i\tau_2$, imposes the constraints $\chi^{(3)}_{1,xxx}=\chi^{(3)}_{2,yyy}$ and $\chi^{(3)}_{1,yyx}=\chi^{(3)}_{2,xxy}$.
Under complex combinations, we organize the cubic response as
\begin{eqnarray}
D^{(3)}_1-iD^{(3)}_2
&=& \chi^{(3)} (B_x+iB_y)^3 
\nonumber \\
&+&  \tilde{\chi}^{(3)} B^2
(B_x-i B_y),
\end{eqnarray}
where $\chi^{(3)}=\frac{1}{4}(\chi_{1,xxx}^{(3)}-3\chi_{2,xxy}^{(3)})$,
 $\tilde{\chi}^{(3)}=\frac{3}{4}(\chi_{1,xxx}^{(3)}+\chi_{2,xxy}^{(3)})$.
The harmonics of the critical current distribution are expressed as \begin{eqnarray}
\Delta J_c^{(3)}&=&
\frac{2}{\pi}
B^3 [\chi^{(3)}
\cos(3\phi +3\phi_B)\nonumber\\
&&~~~~~~+\tilde{\chi}^{(3)}
\cos (3\phi-\phi_B)].
\label{eq:3rd_harm4}
\end{eqnarray}

Similarly, the $S_4$ symmetry operation leads to the constraints
$\chi^{(1)}_{1,xxx}=-\chi^{(1)}_{2,yyy}$ and $\chi^{(1)}_{1,yyx}=-\chi^{(1)}_{2,xxy}$.
Hence, under complex combinations, the cubic response is simplified
\begin{eqnarray}
D^{(1)}_1-iD^{(1)}_2
&=& \chi^{(1)} B^2 (B_x+i B_y)
\nonumber \\
&+& \tilde{\chi}^{(1)} (B_x-iB_y)^3, 
\end{eqnarray}
where $\chi^{(1)}=\frac{3}{4}(\chi_{1,xxx}^{(1)}-\chi_{2,xxy}^{(1)})$,
 $\tilde{\chi}^{(1)}=\frac{1}{4}(\chi_{1,xxx}^{(1)} +3\chi_{2,xxy}^{(1)})$.
Then we arrive at the
harmonics of the critical current distribution as 
\begin{eqnarray}
\Delta J_c^{(1)}&=&
\frac{2}{\pi}
B^3 [\chi^{(1)}
\cos(\phi +\phi_B)\nonumber \\
&&~~~~~~+\tilde{\chi}^{(1)}
\cos (\phi-3\phi_B)
].
\label{eq:3rd_harm5}
\end{eqnarray}



\onecolumngrid

\end{document}